\documentclass[12pt,a4paper]{article}
\usepackage{amssymb,amsmath,amscd,epsfig}
\usepackage[dvips]{color}
\newcommand{\be}{\begin{eqnarray}}
\newcommand{\ee}{\end{eqnarray}}
\newcommand{\bi}{\bibitem}

\newcommand{\rar}{\rightarrow}

\newcommand{\rhoij}{\rho_{ij}}
\definecolor{gold}{rgb}{0.89,0.78,0}
\definecolor{grn05}{rgb}{0,0.5,0}

\begin{document}

{
\title{
{
{
COSMOLOGICAL MAGNETIC FIELDS AND CMBR POLARIZATION}}}
}

\author{
{A.D. Dolgov}
\\[5mm]
{\small\it INFN, Ferrara 40100, Italy}\\
{\small\it ITEP, 117218, Moscow, Russia}
}

\maketitle

\bigskip

\begin{abstract}

A simple introduction to physics of CMBR polarization and the Faraday rotation
of the latter in cosmic magnetic field is presented. The content of the 
lecture is the following:
1. Description of polarization of photons. 
2. Polarization field of CMBR.
3. Faraday effect.
4. Cosmic magnetic fields.
5. Faraday rotation of CMBR polarization.

\end{abstract}
\newpage

\section{Description of polarization of photons
\label{s-pol-gamma}}

The polarization state of an electromagnetic wave is determined by 
the vector of its electric field which is orthogonal to the direction
of the propagation. 
The polarization density matrix is defined as the average value of 
bilinear combinations of the components of the electric field of the
propagating wave~\cite{ll}:
\be
\rho_{ij} =\langle E_i E^*_j \rangle 
\label{rho-ij}
\ee
Evidently,
$\rhoij$ is the 2nd rank tensor in 2-dimensional
$(x,y)$-space if photon propagates along $z$. This tensor has the following 
well known invariants:
\\
1. Trace, which is equal to intensity of radiation: 
\be
T  = \delta_{ij}\rhoij = |E_x|^2 + |E_y|^2
\label{T}
\ee
2. Helicity: 
\be
V = \epsilon_{ij} \rhoij
\label{V}
\label{hel}
\ee
Non-zero helicity $V$ corresponds to circularly polarized photons
when the average value of the photon spin projection on the direction
of the photon propagation is non-vanishing.

As any $2\times 2$-matrix $\rho_{ij}$ can be decomposed in terms of the
unit matrix, $I$ and three Pauli matrices, $\sigma_k$, $k=1,2,3$.
The coefficients $\xi_k$
in front of the latter are called the Stokes parameters:
\be
\rhoij = T \left( {I}/{2} + \xi_k \sigma_k \right)
\label{stokes}
\ee

For a particular and practically interesting case of photons with
a fixed frequency the components of electric fields can be written
as:
\be
E_x = E_0 e_x \exp[{-i\omega t + i\beta_x}] \nonumber \\
E_y = E_0 e_y \exp[{-i\omega t + i\beta_y}],
\label{ex-ey}
\ee
where the components $\epsilon_k$ of the photon polarization vector
satisfy the normalization condition $e_x^2 + e_y^2 = 1$.

The Stokes parameters for this case can be expressed as:
\be
\xi_2 =e_x e_y \sin(\beta_x -\beta_y)
\label{xi-2}
\ee
where $\xi_2$ is {invariant} with respect to rotations and pseudoscalar
with respect to mirror reflection. It
describes circular polarization, i.e. the photon helicity:
\be
\lambda= {\bf s k}/\omega
\label{lambda}
\nonumber
\ee
The other two Stokes parameters 
$\xi_1$ and $\xi_3$ describe linear polarization: 
\be
\xi_3 &=& \left(e_x^2-e_y^2\right)/2 \nonumber\\
\xi_1 &=& e_x e_y \cos (\beta_x -\beta_y)
\label{xi-1-3}
\ee
They transform under coordinate rotation in $(x,y)$-plane
by angle $\phi$ as:
\be
\xi_1' &=& \xi_1\,\cos 2\phi  - \xi_3\,\sin 2\phi  \nonumber\\
\xi_3' &=& \xi_1\,\sin 2\phi  + \xi_3\, \cos 2\phi \xi_3 
\label{xi'}
\ee
Making proper rotation one can always arrange vanishing of one of these
Stokes parameters, e.g.
\be
\xi_1 = 0
\label{xi1-0}
\ee

Using the transformation law (\ref{xi'}) one can check that the following
combinations of the Stokes parameters are the
eigen-vectors of rotation:
\be
\xi_3 \pm i\xi_1 \rar \exp\left[\pm 2i \phi\right] 
\left(\xi_3 \pm i\xi_1 \right)
\label{xi3-ixi1}
\ee

\section {Polarization by Thomson scattering \label{s-thomson}}

As is well known, see e.g. the book~\cite{ll}, elastic scattering of
unpolarized photons on unpolarized electrons
\be
\gamma + e \rar \gamma' + e'
\label{gamma-e}
\ee
produces polarized photons. The reaction amplitude can be written as
\be
A = e'_i {\cal A}_i,
\label{A}
\ee
where $e'_i$ is the polarization vector of photons in the final state.
The polarization matrix of the scattered photons, up to normalization
factor, is expressed as:
\be
\rhoij \sim {\cal A}_i {\cal A}_j^*
\label{ai-aj}
\ee
If we choose the coordinate $z$ in the direction of $\gamma'$ and $x$ 
in the reaction plane and denote by 
$\theta$ the scattering angle then the only non-zero Stokes 
parameter would be equal to
\be
\xi_3 = \frac{\sin^2\theta}{\omega/\omega' + \omega'/\omega -\sin^2\theta}
\approx \frac{\sin^2\theta}{1+\cos^2\theta}
\label{xi3}
\ee
where $\omega$ and $\omega'$ are the energies of the initial and final
photons respectively and the approximation of non-relativistic
electrons is made.

The result (\ref{xi3}) can be easily understood.
The only non-vanishing combination involving ${\bf e'}$ in the amplitude
is:
\be
{\bf e'\, k} \sim \sin\theta
\label{e'k}
\ee
Hence, from eq. (\ref{ai-aj}) follows that
\be
\xi_3 \sim (\sin\theta)^2
\label{xi-3T}
\ee
The other Stokes parameter, which may be non-zero, vanishes
by the choice of coordinate direction, $\xi_1 = 0$, while
$\xi_2 = 0$ due to {\it parity conservation}.

The degree of polarization of the scattered radiation is proportional to
the differential Thomson cross-section, which is equal to~\cite{ll}:
\be
\frac{d\sigma}{d\Omega} = \frac{3}{16\pi}\, \frac{8\pi \alpha^2}{3 m_e^2}\,
(1+\cos^2\theta)
\label{dsigma}
\ee

\section{Polarization of CMBR \label{pol-cmbr}}

According to the previous section
the photons of CMBR could become polarized due to Thomson scattering 
on the electrons in cosmic plasma. However, the polarization must
vanish in {\it homogeneous and isotropic } world. It is evidently
true, simply because there is no preferred direction in such a world.
In realistic cosmology small but non-zero density perturbations
existed in the primeval plasma prior to hydrogen recombination. As we
see below, the polarization of CMBR would be induced by the 
the inhomogeneities in the distribution of photons, i.e. by 
the angular fluctuations of their temperature. 
This effect was noticed many years
ago, the pioneering papers include refs.~\cite{pion}.  
For review and history see e.g.~\cite{rev}.

We assume that circular polarization vanishes. As we mentioned above,
see also ref.~\cite{ddnn-2}, such polarization cannot be induced
due to parity conservation. Bounds on possible new interactions 
violating parity from the absence of circular polarization of CMB
are discussed in ref.~\cite{kam-par}. With the
present day accuracy they are very far from the usual weak interactions.

As we have seen above,
in the usual case of the linearly polarized CMBR radiation field the 
intensity of polarization is described by two functions,
$Q=T\xi_3$ and $U=T\xi_1$, where $T$ is the total intensity of radiation 
with frequency $\omega$:
\be
\bar\rho =
\left[ \begin{array}{cc}
Q & U \\
-U & Q 
\end{array}\right]
\nonumber
\label{rho-bar}
\ee

The total polarization originated 
as a result of the Thomson scattering should be 
obtained by the integration of expression (\ref{xi3}) with account of
(\ref{dsigma}) over all angles $d\Omega = d\cos\theta d\phi$ 
with rotation around $z$ to the common coordinate system:
\be
Q -i U =\frac{\sigma_T}{\sigma_N}\,\int d\omega \sin^2\theta \exp[2i\phi]\,
T\left(\theta,\phi \right)
\label{pol-tot}
\ee
where $\sigma_N$ is some normalization area over which the 
data are averaged.
Thus one can see that the polarization of the scattered radiation is 
proportional to quadrupole moment of the incoming radiation.

\section{Properties of CMBR polarization field \label{s-proprt}}

For description of the CMBR polarization is convenient to know that in
addition to the invariants (\ref{T}) and (\ref{hel}) there exist two more
invariants which can be constructed by differentiating of the 
polarization matrix. One of them is a scalar:
\be
S=\partial_i \partial_j \rho_{ij},
\label{S}
\ee
while the other is a pseudoscalar:
\be
P=\epsilon_{ik}\partial_i \partial_j \rho_{jk}
\label{P}
\ee

If density perturbations are created by a
purely scalar field, then the only way to 
construct a vector out of this scalar is to apply the operator of
differentiation, i.e. gradient, $\partial_j$. A second rank tensor can 
be either obtained by the second order derivatives, 
$\partial_j \,\partial_i$ or by multiplication by the invariant Kronecker
tensor $\delta_{ij}$. In particular, the traceless part of the
polarization tensor should have the form:
\be
\rhoij = \left(2 \partial_i \partial_j -\delta_{ij} \partial^2\right)\Psi
\ee
where $\Psi$ is a scalar function of coordinates.

One can easily check that for the case of scalar perturbations the 
pseudoscalar invariant (\ref{P}) vanishes:
\be
{ P=0 }
\label{P0}
\ee
Thus we arrive to an important conclusion that a
non-zero $P$ is an indication for tensor perturbations
or, more generally, for something beyond scalar perturbations.
A short list of those includes:
\begin{enumerate}

\item{}
Vector perturbations, e.g. magnetic fields. The relevant part of the
polarization matrix can be written in the form:
\be
\rho_{ij} = \partial_i V_j + \partial_j V_i
\label{rho-vec}
\ee
where ${\bf V}$ describes a vector perturbation field. 
Correspondingly,
$P$ may be non-vanishing and equal to:
\be
P_V = \epsilon_{ij}\partial^2 \partial_i V_j
\label{P-vec}
\ee

\item{}
Tensor perturbations, which might be produced by gravitational waves. In
this case the CMBR polarization matrix may contain the contribution: 
\be
\rho_{ij} = \partial_i h_{3j} + \partial_j h_{3i}
\label{rho-tens}
\ee
and the pseudoscalar invariant takes the form:
\be
P_T= \epsilon_{ij} \partial^2 \partial_i h_{3j}
\label{P-tens}
\ee

\item{} 
The pseudoscalar invariant may be non-zero even in the case of purely
scalar perturbations but in the second or higher 
orders~\cite{zb-ad,matr1}. The polarization matrix would be
proportional to the products of gradients of two different scalar
functions $\Psi_1$ and $\Psi_2$:
\be
\rho_{ij} = \partial_i \Psi_1 \partial_j \Psi_2 +
\partial_i \Psi_2\ \partial_j \Psi_1, 
\label{rho-2}
\ee
where e.g. $\Psi_2 =\partial_t \Psi_1$. The $P$-invariant is equal to:
\be
P_2= \epsilon_{ij} 
\left( \partial_i \partial^2 \Psi_1 \,\partial_j \Psi_2 +
 \partial_i \partial^2 \Psi_2 \,\partial_j \Psi_1 \right)
\label{P2}
\ee 
\end{enumerate}

The decomposition of the polarization matrix according to the 
magnitude of the scalars $S$ and $P$ is considered in 
ref.~\cite{s-p}. It is analogous to magnetic/electric or 
B/E - decomposition~\cite{b-e} but the former looks more natural
because no vector field of electric or magnetic type is associated
with $2\times 2$ polarization matrix. 

However, there may be other, non-primordial, sources for 
non-zero $P$ due to propagation of CMB photons through the large
scale structure. They are analyzed in ref.~\cite{non-0-P}.

\section{Eigenvectors of polarization matrix \label{s-eigen}}

In some works polarization fields is characterized by the ``vector'' 
\be
{\bf W} = (Q,U)
\nonumber
\ee
However, ${\bf W}$ is not a vector but some mixture of components
of a second rank tensor. Though such a description is possible, it is 
not convenient~\cite{nas-nov}. 
Field lines of such ``vector'' change under rotation
in a non-trivial way. We see portraits of famous astronomers  and the
portrait of Louis XIV in this beautiful room. 
If someone takes one or other portrait and
rotates it, or ever simpler, just turns one's own head, the 
picture would surely remain the same. This would not
be true for the pictures of CMBR polarization field presented in terms
of ``vector'' ${\bf W}$. 

Much more convenient is to present polarization
maps in terms of field lines of the eigenvectors ${\bf N}$
of the polarization
matrix $\rho_{ij}$~\cite{ddnn-1,ddnn-2}. An interesting feature of such
maps is unusual singularity points of the field lines of the eigenvectors
${\bf N}$. As is known from the classical analytical mechanics,
dynamical systems may normally have the following three 
singularity types:
saddle, focus, and knot. However, the eigenvectors ${\bf N}$ are not
analytic near singularity, where $|{\bf N}|=0$. Because of that the 
character of singularities changes. There still remain three types
of them but now they look differently, see fig. 1. We suggested the 
following names for them: saddle (the same as above but now it has
three, instead of four, straight line asymptotic), beak, and 
comet~\cite{ddnn-1,ddnn-2}. An analysis of statistical properties
of these singularities is performed in the 
papers~\cite{ddnn-2,nas-nov,ddnn-1,va-lu}.

In fig. 2 the simulated map of CMBR polarization is presented,
according to ref.~\cite{ddnn-2}. One can easily see the singular
points corresponding to the three types indicated in fig. 1. 
In contrast to description of the direction of the polarization field
in terms of ${\bf W}$, the map of field lines of ${\bf N}$ remains
the same independently of the angle at which one observes it. On the
other hand, as we mentioned above, different singular points of the
map of the field lines of ${\bf W}$  transform into each other under
rotation and the general picture becomes completely different.

\begin{figure}
\centering
\epsfxsize=15cm
\epsfbox{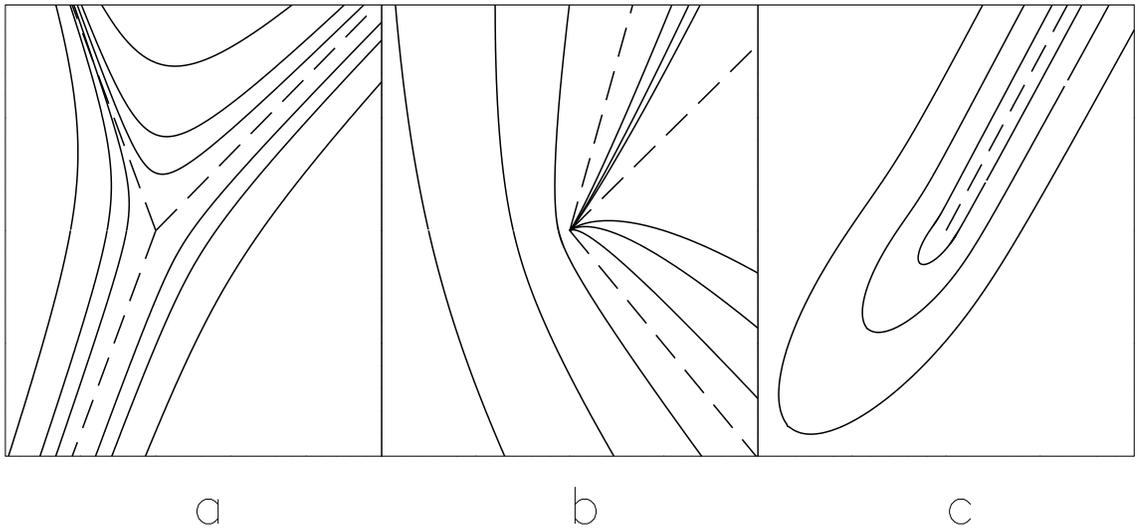}
\vspace{0.75cm}
\caption{
Flux lines for three different
types of singular points:
(a) saddle, (b) beak, and (c) comet.
Dashed lines show peculiar solutions, separatrix.
}
\end{figure}

\begin{figure}
\centering
\epsfxsize=15 cm
\epsfbox{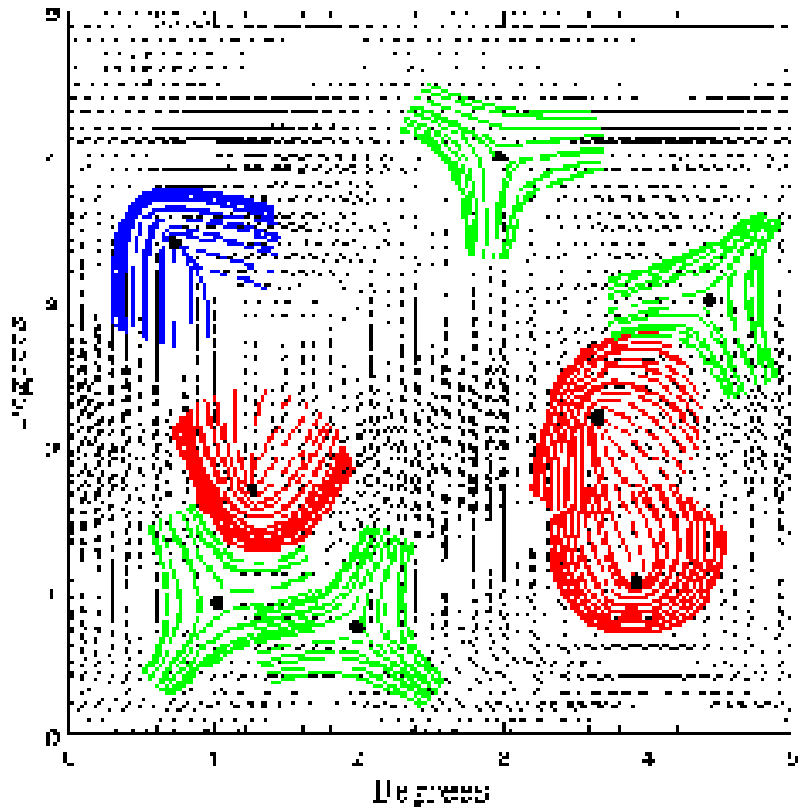}
\vspace{0.5cm}
\caption{Simulated map of CMB polarization eigen-vector field ${\bf N}$
Solid lines show the flux line behavior near
singular points where polarization vanishes.
}
\end{figure}

\section{Faraday effect \label{far-eff}}

The Faraday effect is a rotation of the polarization plane of linearly 
polarized photons in magnetic field ${\bf B}$, due to interaction 
with electrons in the medium. Such rotation is generated by the
breaking of the reflection symmetry in presence of magnetic
field. In a medium without reflection symmetry refraction indices 
for left- and right-handed photons are different, $n_+ \neq n_-$.
Linearly polarized electromagnetic 
wave can be decomposed into two rotationally
polarized ones rotating in opposite directions. We assume that 
the electric field of the initial wave was directed along axis $x$.
Then using the simple identity: 
\be
1 = (1+i)/2 + (1-i)/2
\label{1i}
\ee
we expand $E_x$ in terms of two helicity eigenstates of the incoming
photon:
\be
E^{(in)} = E_x = (E_+ + E_-)/2
\label{1}
\ee
Each helicity state propagates independently and acquires phase 
proportional to the distance of propagation:
\be
E_{\pm}^{(fin)} = \exp \left(i k_\pm l \right)\, E_{\pm}^{(in)} 
\label{E-pm}
\ee
Since the refraction index for different helicity states is different
by assumption, then $k_\pm = k_0  \pm \Delta k$ and 
\be
E_{x}^{(fin)} = E_{x}^{(in)} \exp [i k_0 l ] \cos (\Delta k l)\nonumber\\
E_{y}^{(fin)} = E_{x}^{(in)} \exp [i k_0 l ] \sin (\Delta k l)
\nonumber
\label{E-fin}
\ee
The relative phase remains zero. It means that the initially linearly
polarized wave remains linearly polarized but the polarization plane  
is rotated by the {\it rotation angle}: 
\be
\Phi = \arctan \left[ E_{y}^{(fin)}/E_{x}^{(fin)} \right] = \Delta k l 
\nonumber
\label{Phi}
\ee

Thus the calculation of rotation angle is reduced to the calculation of
the refraction index of ionized gas - this is the cosmic medium 
where photons of CMBR propagate. 
The equation of motion of electrons, with charge $(-e)$,
in superimposed external (large) magnetic field ${\bf B_0}$ and
(weak) electromagnetic wave ${\bf E}\exp(i\omega t)$ reads:
\be
\ddot{\bf r} = e{\bf B_0}\times \dot{\bf r} - e{\bf E} \exp[i\omega t]
\label{ddot-r}
\ee
To solve this equation we need to 
decompose the propagating wave in terms of helicity states: 
\be
{\bf E} = C_+\left({\bf n}_x +i {\bf n}_y\right) +
C_- \left({\bf n}_x -i {\bf n}_y\right)
\label{dcmps}
\ee
for which the equation diagonalizes and can be solved as
\be
{\bf x}_\pm = \frac{e{\bf E}_\pm}{m\omega\left(\omega \mp \omega_B
\right)}
\ee
where $\omega_B = eB_0/m$.

The electric polarization moment of the plasma is easily found:
\be
{\bf {\cal P}}_\pm = -N_e e {\bf x}_\pm
\label{P-el}
\ee
and correspondingly the dielectric constant is 
\be
\epsilon_\pm = 1 + 4\pi {\cal P}/E =
1+ \frac{4\pi e^2  N_e}{m\omega \left(\omega \mp \omega_B\right)}
\nonumber
\label{epsilon}
\ee

The plasma refraction index for different helicity states of photons
is given by the standard expression: 
$n_\pm=\sqrt{\epsilon_\pm}$ and thus the
differential Faraday rotation angle is equal to: 
\be
\frac{d\phi}{dl} = \frac{2\pi N_e e^3 B_0}{m^2 \omega^2}
\label{dphi}
\ee
where $m$ is the electron mass, $e^2=\alpha= 1/137$, and 
$N_e$ is the number density of electrons.
Usually the results is presented in terms of frequency 
$\nu =\omega/(2\pi)$ or wave length $\lambda = 1/\nu$.

\section{Cosmic magnetic fields \label{cosm-magn}}

The origin of large scale cosmic magnetic fields remains one of
cosmological mysteries. They surely exist in galaxies with the
coherence scale of a few kpc and rather large field 
strength: 
\be
B_{gal} = {\rm a\,\,\, few\,\,\,} \mu{\rm G},
\label{B-gal}
\ee
for a review see\cite{magn-obs-rev}, more recent data can be found 
e.g. in ref.~\cite{magn-gal}.
Less certain are the observational data in favor of existence of 
intergalactic magnetic fields at the scales $\sim 1$ Mpc. 
Still there are rather convincing indications in favor of the latter
with the amplitude $B_{ig} \sim 10^{-3} B_{gal}$. 
It is rather interesting that the amplitude of these fields is related
to the galactic field by the inverse ratio of the corresponding
scales squared. This would be so if the adiabatic compression
took place and the field was amplified as
$B\sim 1/l^2$. Since the ratios of scales are
\be
l^{(in)}_{gal}/l_{gal}&\sim& 10^2, \nonumber \\
 l^{(in)}_{ig}/l_{ig}&\sim& 3,
\label{l-ig}
\ee
one would expect $B_{gal}\sim 10^3 B_{ig}$, if galactic and intergalactic
magnetic fields have a common origin and
galactic dynamo amplification was not effective. 
The latter might be quite efficient. According to different 
estimates~\cite{dynamo} the amplification factor could be as large
as $10^{15\pm 5}$. 
If galactic dynamo indeed amplified primordial seed fields by this
large factor, it means that the original fields were too weak to
be observable by the Faraday rotation of the CMBR polarization.
Otherwise, if $B_{ig}\sim 10^{-9}$ Gauss, the effect may be noticeable.

There is rather strong evidence accumulated
recently~\cite{cluster} in favor of quite strong magnetic fields
in galactic clusters with the magnitude close to those in galaxies. 
If this is indeed true, then the chances of observation of the 
Faraday rotation of CMBR polarization would be even higher.

A vast literature exists on possible mechanisms of generation of 
primordial magnetic fields (for a review see e.g.~\cite{magn-rev}).
Different mechanisms can be roughly speaking separated into the 
following classes:
\begin{enumerate}
\item{} Galactic processes, stellar phenomena and reconnection of 
field lines.
\item{} Processes during structure formation.
\item{} Processes during the recombination epoch;
vorticity, $\nabla \times V$, may be generated in the second order in
density perturbations.
\item{} Processes in the early universe: \\
a) inflation, might produce small fields but at large scales;\\
b) phase transitions could create large fields but at small scales.
\end{enumerate}

All these mechanisms, except for the first one, 
might create noticeable fields at 
the CMBR decoupling which can be 
potentially observable by the Faraday rotation of the CMBR
polarization. This effect was first discussed in ref.~\cite{kos-loeb}
and attracted considerable attention~\cite{fr1,cdmv,kos-kah} during
the last years. 

\section{An estimate of the rotation angle \label{s-estim}}

The dominant contribution to the rotation angle comes from the period
near the hydrogen recombination epoch. Indeed, the dependence 
of the rotation angle on the cosmic scale factor is:
\be
d\Phi \sim \lambda^2 N_e B\, a\, d\eta 
\sim a^2\, \frac{1}{a^3}\,\frac{1}{a^2}\,a \sim \frac{1}{a^2},
\label{d-phi}
\ee
where
$\eta$ is the conformal time, the magnetic field is assumed to evolve
as $B\sim 1/a^2$ (adiabatic compression and no dynamo amplification).
Thus the rotation is dominated by an early epoch.
Before the recombination the photon mean free path along which the 
rotation angle could be ``accumulated'',
$l_{free}$, is very small and practically $\langle\Phi\rangle=0$.
After, the recombination the number density of free electrons, $N_e$,
drops down by the factor $10^{-5}$. Thus the most favorable period
for the rotation of the 
polarization plane in large scale magnetic fields
is the recombination epoch.

The differential rotation angle is equal to:
\be
\frac{d\Phi}{d\eta} = \frac{x_e N_e e^3 a}{2\pi m^2 \nu^2}\,{\bf B n}
\label{dphi/deta}
\ee
where $x_e$ is ionization fraction and ${\bf n}$ is the unit vector in
the direction of propagation of radiation.
By assumption of adiabatic compression
$Ba^2 = const = B_0 a_0^2$, i.e. it 
is equal to the present day value.

The optical depth is expressed through the Thomson cross-section
$\sigma_T$ and the number density of free electrons and is equal to
\be
\frac{d\tau}{d\eta} = N_e\sigma_T a
\label{opt-d}
\ee
The total rotation angle (for homogeneous field along photon
propagation) is easily estimated:
\be
\Phi = \frac{3\lambda^2_0 {\bf B}_0 \cdot {\bf n}}{16\pi^2 e} 
\int d\tau \exp\left(-\tau\right)=
\frac{3\lambda^2_0 {\bf B}_0 \cdot {\bf n}}{16\pi^2 e} 
\label{phi-tot}
\ee
(here the usual in this field, but unusual for particle physics,
convention $e^2=\alpha$ is used).

Numerically we find:
\be
\Phi \approx
{ 2^o \left( \frac{B_0}{10^{-9}{\rm {Gauss}}}\right)\,
\left(\frac{30\,{\rm {GHz}}}{\nu_0}\right)^2
}
\label{numeric}
\ee
(it helps to know that 1 Gauss = $6.9\cdot 10^{-14}$ MeV$^2$).

\section{Statistical properties of magnetic fields and rotation
angle \label{s-stat}}

We express all the quantities below in terms of the present day values,
in particular:
\be
{\bf B}_0 ({\bf x}) = a^2(\eta)\,{\bf B} ({\bf x}, \eta).
\label{B0-B}
\ee

As usually the field amplitude is expanded in terms of its
Fourier modes:
\be
{\bf B}_0 ({\bf x}) = \frac{1}{(2\pi)^3} \int d^3 k e^{-i {\bf k x}}\,
{\bf b_0}({\bf k})
\nonumber
\label{b-B}
\ee
Below we omit sub-zero for simplicity of notations.

The correlator of the field amplitudes is:
\be
\langle B_i ({\bf x_1}) B_j ({\bf x_2})\rangle=
C_{ij} (|\bf x_1 -\bf x_2 |) 
\label{Cij}
\ee
The correlator depends on the difference of the coordinates
because of the average homogeneity and isotropy.
Correspondingly the Fourier modes should be delta-correlated.
Their correlator can be expressed through two scalar functions
$A(k)$ and $S(k)$~\cite{caprini}:
\be
\Pi_{ij}= 
\langle b_i ({\bf k_1}) b_j^* ({\bf k_2})\rangle=
2\,(2\pi)^3\delta \left({\bf k}_1-{\bf k}_2\right) \nonumber \\
\left[ \left( \delta_{ij} - \kappa_i\kappa_j \right)\, S(k) +
i\epsilon_{ijl} \kappa_l A(k) \right] 
\label{Pij}
\ee
where $\kappa_i = k_i/|{\bf k}|$.

The energy of magnetic field depends only on $S(k)$:
\be
\int d^3 x B_j^2 = \frac{2}{\pi^2}\int dk k^2 S(k)
\label{dx-B} 
\ee
while $A(k)$ determines the so called helicity of magnetic 
field~\cite{hel}.

The correlator of the rotation angle can be expressed through the
correlator of magnetic field as:
\be
\langle \Phi ({\bf n}) \Phi ({\bf m})\rangle =
\left(\frac{3}{16 \pi^2 e}\right)^{2}
\int \, d\eta \, g(\eta) \,\,\,\,\,\,\,\,\,\,\,\,\,\,
\nonumber \\
\int \, d\eta' g(\eta') \langle
\left[{\bf B}_{0} (\Delta \eta \;{\bf n}) \cdot {\bf n}) ({\bf B}_{0}
(\Delta \eta^\prime \;{\bf m})\cdot{\bf m}\right] \rangle 
\nonumber
\label{Phi-Phi}
\ee
where $g(\eta) = (d\tau/d\eta) \exp [-\tau (\eta)]$
and $\delta \eta = \eta - \eta_0$.

Using eq. (\ref{Pij}) one finds
\be
\langle ({\bf B}\cdot {\bf n}) ({\bf B}\cdot {\bf m}) \rangle  = 
\frac{1}{2(2\pi)^3} \int
d^{3}{k}  \left\{ \left[  ( {\bf n} {\bf m} )
      - ({\bf n} \cdot {\bf \kappa}) ({\bf m} \cdot{\bf \kappa}) \right]
 S(k) \right. + 
\nonumber \\
\left. i \left[({\bf n} \times {\bf m})
\cdot {\bf \kappa}\right] A(k) \right\} 
\exp \left[ -i{\bf k}
\left( {{\bf n}} \Delta \eta - {{\bf m}} \Delta \eta'
\right) \right]  
\hspace{2.8cm}
\label{B0B0}
\ee
It can be checked that the term containing $A$ 
vanishes~\cite{cdmv,kos-kah}. It prevents from measurements of helicity
of magnetic field through Faraday rotation, in contrast to the
results of earlier papers~\cite{pog}.

Performing the necessary integration we obtain:
\be
\langle ({\bf B} {\bf n}) ({\bf B}{\bf m}) \rangle = 
\left[  ({\bf n}{\bf m} ) C_{\perp}(r)
+ ({\bf n} {{\bf r}}/{r}) ({\bf m}{{\bf r}}/{r})
        (C_{\parallel}(r)-C_{\perp}(r)) \right], 
\label{int-B}
\ee
where ${\bf r}={{\bf n}} \Delta \eta - {{\bf m}} \Delta \eta'$, and
\be
C_{\perp}(r) = 
\frac{2}{3(2\pi)^3}\int_{0}^{\infty} dk 
{\cal E}_{\rm B}(k) \left[ j_{0}(kr) -  \frac{j_{2}(kr)}{2} \right ] 
\nonumber \\
 C_{\parallel}(r) = 
\frac{2}{3(2\pi)^3}\int_{0}^{\infty} dk {\cal E}_{\rm B}(k)
\left[ j_{0}(kr) +  j_{2}(kr) \right],
\ee
Here $j_{i}(x)$ are the spherical Bessel functions of the
$i^{th}$ order and ${\cal E}_{\rm B}(k)$ is the magnetic power
spectrum:
\be
E_B = \int^infty_0 dk {\cal E}_{\rm B}(k),
\nonumber \\
{\cal E}_{\rm B}(k) = 2\pi k^2 S (k)
\label{E-of-k}
\ee
The function $S(k)$ is usually parametrized as
\be 
S(k) = S_0 k^{n_S} \exp [-(k/K)^2]
\label{S-of-k}
\ee
For this simple form of $S(k)$ the integral can be taken analytically.
The correlation function of the Faraday rotation measure, according
to ref.~\cite{cdmv} is presented in fig. 3. The results are not
particularly sensitive to the power $n_S$.

\begin{figure}[htb]
\epsfbox{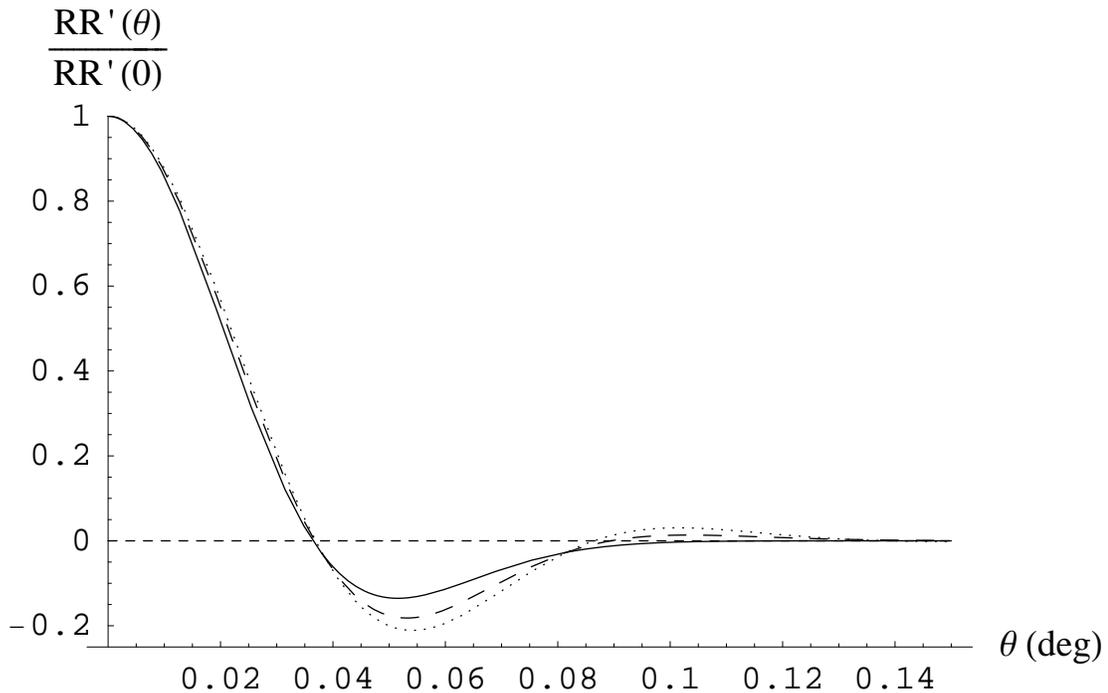}
\caption{Faraday rotation measure
correlation $RR'(\theta)$ as a function of the separation angle
$\theta$. The three lines correspond to the magnetic field
spectral index $n_S =2$ ({\it solid line}), $n_S = 4$ ({\it dashed
line}) and $n_S = 6$ ({\it dotted line}). The correlation length
of the magnetic field is taken as $\xi = 20$ Mpc.}
\end{figure}

\section{Conclusion \label{s-concl}}

\begin{enumerate}
\item{}
We do not understand how large scale cosmic magnetic fields 
have been formed. If $B_{gal}$ and $B_{ig}$ have the same origin 
and galactic dynamo did not operate, impact of primordial 
fields could be observable in CMBR polarization.
Another way around, an observation of the Faraday rotation of
the CMBR polarization can bring an important information about
primordial magnetic fields and help to solve the problem of their
origin. 
\item{}
$P$-type (or $B$-type) polarization created by magnetic fields
may mimic gravitational waves 
but they could be possibly distinguished due to different
frequency dependence.
\item{}
Eigenvector description of the polarization field may be useful.
Their statistical properties may depend upon the strength
of the primordial magnetic fields. This problem deserves further 
consideration.
\end{enumerate}

\end{document}